\documentclass{emulateapj}
\usepackage{natbib}
\usepackage{xspace}
\usepackage{amssymb}
\usepackage{amsmath}
\usepackage{graphicx}

\def\aap{A\& A}

\def\apj{ApJ}

\def\apjl{ApJL}
\def\apjs{ApJS}

\def\rcentr{R_{\mathrm{centr}}\xspace}
\def\rcentrt{R_{\mathrm{centr}}(t)\xspace}

\def\dminf{\dot M_{\mathrm{infall}}\xspace}
\def\mcore{M_{\mathrm{core}}\xspace}
\def\rcore{R_{\mathrm{core}}\xspace}
\def\bcs{{\bar c}_s\xspace}
\def\mdisk{M_{\mathrm{disk}}\xspace}
\def\mstar{M_{*}\xspace}
\newcommand{\Msun} {M$_\odot$}

\newcommand{\macc} {$\dot M_{\mathrm{acc}}$}
\newcommand{\Ha} {H$\alpha$}
\newcommand{\Roph} {$\rho$~Oph}
\newcommand{\simless}{\mathbin{\lower 3pt\hbox
      {$\rlap{\raise 5pt\hbox{$\char'074$}}\mathchar"7218$}}} 
\newcommand{\simgreat}{\mathbin{\lower 3pt\hbox
     {$\rlap{\raise 5pt\hbox{$\char'076$}}\mathchar"7218$}}} 

\begin{document}

\shorttitle{}
\shortauthors{Dullemond, Natta and Testi}
\title{Accretion in protoplanetary disks: the imprint of core properties}
\author{C.P.~Dullemond}
\affil{Max Planck Institut f\"ur Astronomie, K\"onigstuhl 17,
D69117 Heidelberg, Germany; Email: dullemon@mpia.de}
\author{A.~Natta and L.~Testi}
\affil{Osservatorio Astrofisico di Arcetri,
Largo E.~Fermi 5, 50125 Firenze, Italy}

\begin{abstract}
In this Letter we present a theoretical scenario to explain the steep
correlation between disk accretion rates and stellar masses observed in
pre-main sequence stars. We show that the correlations and spread observed
in the two best studied regions, \Roph\ and Taurus, can be reproduced by a
simple model of single star formation from a rotating collapsing
core and the viscous evolution of the circumstellar disk. In this model, the
rate of rotation of the parent core sets the value of the `centrifugal
radius' within which the infalling matter is loaded onto the surface of the
disk. As a consequence, the disk accretion rate measured long after the
dispersal of the parental core bears the imprint of the initial conditions
of star formation. The observed trend results naturally if, at the onset of
the collapse, cores of all masses rotate with the same distribution of
angular velocities, measured in units of the break-up rotation rate.
\end{abstract}


\section{Introduction}
Young stars are surrounded by disks for a large fraction of their pre-main
sequence evolution. The accretion of material from such disks controls the
disk properties and its evolution with time. Measurements of the mass
accretion rate onto the star (\macc) were obtained for T Tauri stars (TTS)
in Taurus  several years ago, showing a large spread of \macc\ for
stars of  similar properties (Gullbring et al.~\citeyear{Gea98}). 

The discovery of disks around very low mass stars and brown dwarfs (BDs) and
the measurements of the accretion rate in these disks has provided an
unexpected result, namely that, underlying the known large spread, there is
a steep dependence of \macc\ on the mass of the central object, roughly as
\macc$\propto M^{1.8}$ (Muzerolle et al.~\citeyear{Mea03}; Natta et
al.~\citeyear{Nea04}). This correlation only became clear when a
sufficiently large interval of stellar mass $M_{*}$ (roughly two orders of
magnitude) was accessible to observations.

The physical origin of this correlation is unknown.  There has been a number
of suggestions that it may be due to mass-dependent stellar properties, such
as its X-ray luminosity, which may affect the disk physical conditions and
the angular momentum transfer (e.g., Muzerolle et al.~\citeyear{Mea03};
Mohanty et al.~\citeyear{Subu05}; Natta et al.~\citeyear{Nea06a}).
\cite{Pea05} proposed a scenario of Bondi-Hoyle accretion to explain
it. \cite{AA06} showed that the observations can be accounted for by disk
viscous evolution if the ratio of the disk to stellar mass at the end of the
core collapse is fixed, but the viscous time scale is proportional to
1/$\mstar$; they then interpret the spread of \macc\ values as an age
effect.

In this Letter, we propose a different, somewhat more basic scenario. We
follow the collapse of a rotating molecular cloud core and the simultaneous
viscous evolution of the resulting disk. Assuming that cores of vastly
different mass have similar rotation rates measured in units of their
breakup rate, a trend of \macc $\propto M_{*}^{1.8\pm 0.2}$ is naturally
reproduced. The width of the correlation can, in this very simple model, be
traced back to a spread in core rotation rates. In this way, the \macc
measured in the pre-main-sequence phase bears an imprint of the original
core properties, as was already discussed qualitatively by \citet{Nea06a}.

\section{Data}\label{sec-data}

Disk mass accretion rates have been measured in many objects in star forming
regions. In two regions, Taurus and \Roph, they are known for a large number
of objects, ranging in mass from few solar mass to brown dwarfs. The \Roph\
sample (Natta et al.~\citeyear{Nea06a}) includes almost all the Class II
objects (i.e., objects with circumstellar disks) in the region, selected
from their mid-IR excess; the accretion rate is measured from the luminosity
of the infrared hydrogen recombination lines. The Taurus sample is more
heterogeneus, including classical T Tauri stars for which the mass accretion
rate has been determined from the optical and UV veiling, as well as a
number of optically selected BDs where it has been derived by fitting the
\Ha\ profiles with magnetospheric accretion models (see Muzerolle et
al.~\citeyear{Mea05}, and references therein) or from the Ca IR lines
(Mohanty et al.~\citeyear{Subu05}). The presence of a disk in the BDs is not
always known. A comparison of the results in \Roph\ and Taurus can be found
in Natta et al.~(\citeyear{Nea06a}).  The accretion rates turn out to be very
similar in the two regions; the relation between \macc\ and $\mstar$ has
practically the same slope (\macc$\propto\mstar^{1.8\pm0.2}$) and the range
of values for any $\mstar$ is comparable.  The only difference appears in
the BD regime, where Taurus seems to lack high accretors. Given the
different selection criteria for BDs in the two regions (optical colors
vs. infrared excess), and the uncertainty of the \macc\ estimates in the
BDs, it is presently difficult to evaluate the significance of this.

\section{Model}\label{sec-model}
Our model is quite similar to the model described by Hueso \& Guillot
(\citeyear{huesoguillot:2005}, henceforth HG05; see also Nakamoto \&
Nakagawa \citeyear{1995ApJ...445..330N}). It has previously been used in the
context of a theoretical study of the crystallinity of dust (Dullemond, Apai
\& Walch \citeyear{DAW06}). In this section, we describe briefly the model
for the collapsing cloud and disk evolution, and refer for more details to
Dullemond, Apai \& Walch (\citeyear{DAW06}) and HG05.

\subsection{Collapsing cloud}
We start with a core with temperature $T$, turbulent velocity dispersion
$\Delta v$, mass $\mcore$ and solid-body rotation rate $\Omega$
(radian/second). We assume that the core forms a {\em single} star
surrounded by a disk.  For simplicity we assume the core to be a singular
isothermal sphere (Shu~\citeyear{Shu77}). 
After onset of collapse (at which time we set our clock to $t=0$)
the infall rate is $\dminf=0.975\;\bcs^3/G$ (Shu \citeyear{Shu77}), where
$\bcs$ is the {\em effective} sound speed, which combines the thermal and
turbulent pressure.  In the Shu model used here, the infall rate stays
steady until the mass $\mcore$ has accreted onto the star+disk system, and
then it abruptly stops.

The original core has a solid-body rotation rate $\Omega$, which
is slower than the break-up rotation rate
$\Omega_{\mathrm{break}}=\sqrt{GM_{\mathrm{core}}/\rcore}$, where
$\rcore$ is the radius of the core.  According to the singular isothermal
sphere model, this radius is $\rcore=G\mcore/(2\bcs^2)$, so the
breakup rotation rate can be written as
$\Omega_{\mathrm{break}}=2\sqrt{2}\bcs^3/GM_{\mathrm{core}}$.

Upon collapse the core matter will spin up due to the conservation of
angular momentum, leading to the formation of a disk. The infalling matter
falls onto the disk {\em within} the centrifugal radius $\rcentrt{}$, which
increases with time as $\rcentrt\propto t^3$ until the end of the infall
phase. In the following we will name the final (maximum) $\rcentrt$ to be
{\em the} centrifugal radius of the cloud. This radius is:
\begin{equation}\label{eq-rcentr-in-rcore}
\rcentr{} = \frac{\Omega^2\rcore^4}{G\mcore}
\end{equation}
$\rcentr$ can be expressed in terms of the dimensionless rotation rate
$\omega= \Omega/\Omega_{\mathrm{break}}$:
\begin{equation}\label{eq-rcen-gm}
\rcentr{} = \frac{G\mcore}{2\bcs^2}\;\omega^2
\end{equation}
In our calculations, we will assume that the range of possible values of
$\omega$ is the same for cores of very different mass, or in other
words, that the range in ratios of rotational energy over gravitational
energy is independent of core mass.  This means that small, low mass cores
have lower values of $\Omega$ than big, high mass ones.

An important consequence of this assumption is that, for a given value of
$\omega$ larger mass stars form from cores with larger centrifugal
radii. This will turn out to play an important role in the of the \macc$-
\mstar$ relation that follows from the model, as we will discusse below.

\subsection{Disk formation and evolution}
As matter from the core accretes onto the disk, the disk itself evolves,
expanding outwords.  The surface density $\Sigma(R,t)$ changes with time
while matter continues to fall onto the disk within $\rcentrt$ (see
HG05). The disk evolution continues after the core has disappeared, until it
has entirely accreted onto the star.  
The radial velocity $v_R$ is proportional to the 
viscosity coefficient $\nu\equiv \alpha
kT_m/\mu m_p\Omega_K$. Here $\alpha$ is the parameter of viscosity (Shakura
\& Sunyaev \citeyear{shaksuny:1973}), which we take to be constant, $T_m$ is
the midplane temperature, and $\Omega_K$ is the Keplerian frequency.  The
midplane temperature is determined by taking into account the heating due to
viscosity and to irradiation by the central star and by the accretion shock.
We include an effective viscosity caused by gravitational instabilities in
regions where the Toomre parameter drops below unity.

The evolution with time of the star+disk+core system is discussed in detail
in HG05. At time $t_0$, when the core has accreted completely onto the
star+disk system, the ratio of the disk to the stellar mass is an increasing
function of $\rcentr$; it exact behaviour depends on the interplay of the
infalling matter with the viscous evolution of the disk.  If one forces
$\rcentr$ to be independent of $\mcore$, then $\mdisk (t_0) \propto \mcore$,
as expected; if, as in eq.(2), $\rcentr \propto \mcore$, we find that the
disk mass increases approximately as $\mstar^2$.  The following disk
evolution at time $\gg t_0$ is approximately self-similar (see, e.g.,
Hartmann et al.~\citeyear{Hea98}), so that at any given time $t\gg t_0$ the
accretion rate is to zero order proportional to $\mdisk$.

\section{Results}\label{sec-results}
\subsection{Model parameters}
Since we wish to compare our model results to the Taurus and Ophiuchus
samples we need to choose core properties appropriate for these regions.
The average measured line widths \citep{MB83,BAM01} lead to infall rates of
$\dminf\simeq 10^{-5}\,M_{\odot}/$yr ($\bcs=0.35$ km/s) in Taurus and
$\dminf\simeq 3\times 10^{-5}\,M_{\odot}/$yr ($\bcs=0.5$ km/s) in
Ophiuchus. We shall take these two values as representative for these two
star formation regions. The ages of stars in Ophiuchus and Taurus show a
peaked distribution with median values of $\sim 0.5$~Myr and $\sim
1$~Myr respectively, with a small tail of objects at older ages (Palla and
Stahler \citeyear{PS00}).  With these typical ages and the mass infall rates
chosen above, the star formation time scale is always much shorter than the
age of the system. This is consistent with the lack of correlation between
spectral type and stellar age in both regions.

With $\bcs$ chosen at some representative value, and $\mcore$ varied to
cover a range of masses between $0.03$ and $2.5\;M_{\odot}$, the only
remaining parameters of the model are $\alpha$ and $\omega$. Rotation rates
are measured in a limited number of cores, in general more massive than the
cores considered here.  The measurements are consistent with $\omega$ in the
range 0.03--0.5, but there is a large fraction ($\sim$70\%) of
non-detections (Goodman et al.~\citeyear{Gea93}; Belloche et
al.~\citeyear{BAM01}).  In this Letter, we will vary $\omega$ between a very
low value (0.01) and a rather large one (0.3), and investigate the effect of
this parameter on the \macc -- $\mstar$ correlation.

Finally, we chose for the viscosity parameter $\alpha$ the value 0.01, which
has been argued to fit well the properties of TTS disks (Hartmann et
al.~\citeyear{Hea98}), but it remains a very uncertain parameter
because the physics of the anomalous viscosity of disks is not yet fully
understood.

\subsection{Results for the \macc$ - \mstar$ relation}
\begin{figure*}
\centerline{
\includegraphics[width=8cm]{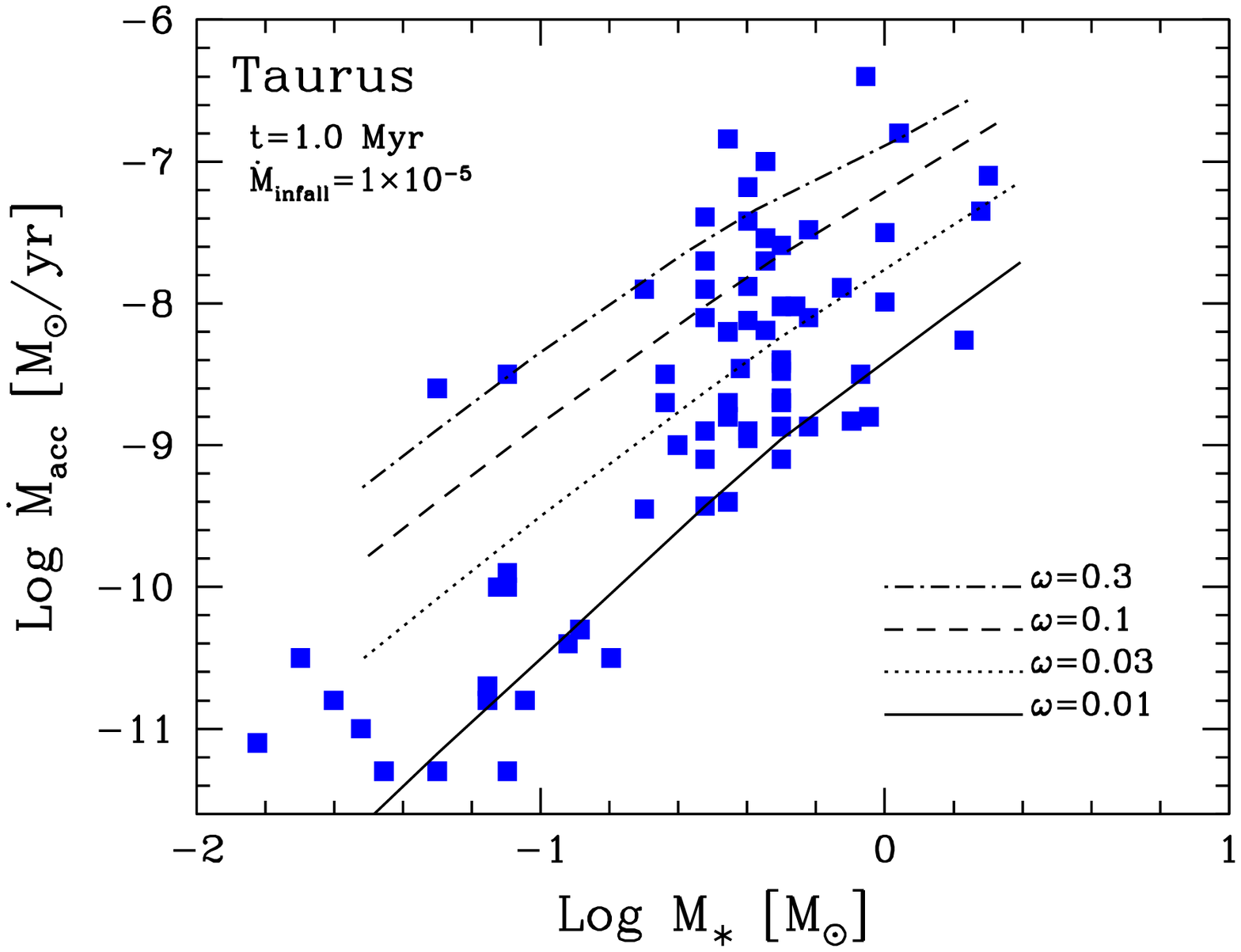}
\includegraphics[width=8cm]{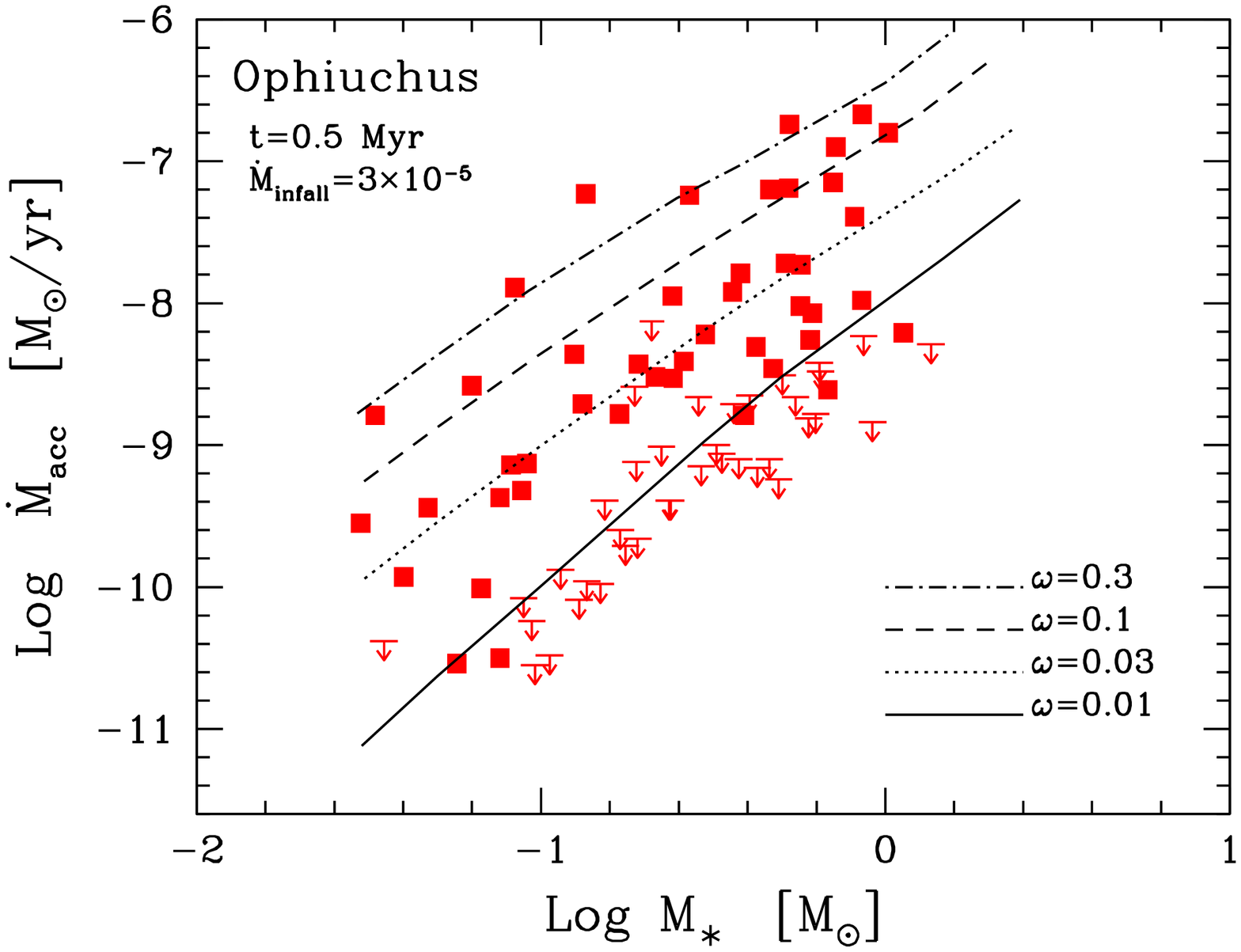}
}
\caption{\label{fig-main} Mass accretion rate \macc\ as function of the mass
of the central object $\mstar$ in Taurus (left panel) and \Roph (right
panel).  Symbols are measured quantities or upper limits.  Each line is a
model series for fixed $\omega$, as labelled, varying $\mcore$. In all model
series $\dminf$ and the snap-shot time are kept constant ($10^{-5}$ \Msun/yr
and 1\,Myr in Taurus and $3\times 10^{-5}$ \Msun/yr and 0.5\,Myr in \Roph).
}
\end{figure*}

Fig.~\ref{fig-main} compares the model results to the observations for
Taurus (left panel) and \Roph\ (right panel).  Each line corresponds to
fixed $\omega$ and fixed $\dminf$ and time, as discussed before.  For each
$\omega$, the models predict a relation between \macc\ and $\mstar$ roughly
as $\mstar^{1.8}$, which is in very good agreement with the trend in the
data.  Lower values of $\omega$ give lower values of \macc\ for fixed
$\mstar$, and we find that values between $0.3$ and $\simless0.01$ can
account for the spread of observed \macc\ at all $\mstar$.

The {\it slope} of the correlation \macc--$\mstar$ is practically
independent of the exact values of any of the model parameters.
However, the {\it value} of
\macc\ for any given $\mstar$ depends on the value of the parameters;
A spread in age, $\dminf$ or $\alpha$ among objects in a region introduces a
spread in the values of \macc\ for a given $\mstar$.  From our model
results, however, we estimate that the observed spread in \macc\ is likely
dominated by the spread in $\omega$. 
\macc\ is not a strong function of
$\dminf$ or $\alpha$ (less than linear). It 
decreases  significantly with time
(rougly as $t^{1.7-1.8}$), but
the observed spread of more than two orders of magnitude (in both regions)
requires that the objects are homogeneously distributed over a factor $\sim
20$ in age; this is much more than observed, since only few objects in each
region have ages so different from the median values (Palla and Stahler
\citeyear{PS00}).

\begin{figure}
\centerline{
\includegraphics[width=8cm]{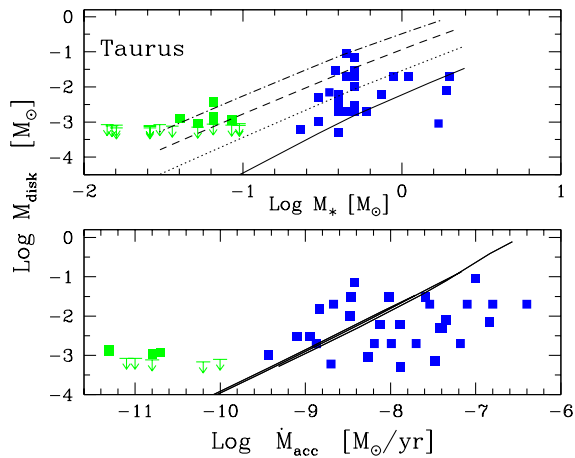}
}
\caption{\label{fig-mdisk} Disk masses as function of $\mstar$ (top
panel) and of \macc\ (bottom panel) for the Taurus sample; binaries
unresolved in the millimeter observations have not been included. Only a few
of the Taurus BDs shown in the top Panel have measured \macc\ and appear in
Fig.~\ref{fig-main} and in the bottom panel. Lines as in Fig.~1.
Note that all models practically overlap in the bottom panel.}
\end{figure}

\section{Discussion}\label{sec-discussion}
The models presented in this Letter account very well for the observed
correlation between \macc\ and $\mstar$ in both star-forming regions.  Note
that the model parameters have not been adjusted to fit the data, but are
chosen {\it a priori} from independent observational evidence.

The general agreement with the slope \macc$\propto\mstar^{1.8}$ is recovered
for all values of the infall rate, age, etc., as long as $\omega$ is the
same for cores of all masses. This is a direct consequence of the dependence
of the centrifugal radius on $\mcore$ (eq.5), as the accretion rate in the
disk depends on the mass reservoir in the outer disk.  If we assume
$\rcentr$ independent of $\mcore$, we recover the well known result that
\macc\ is roughly proportional to $\mstar$.

The scenario we propose has a number of implications, which can be
tested. Firstly, we assume that the same formation process holds for objects
differing in mass by two orders of magnitude, from intermediate-mass stars
to BDs. If so, cores capable of forming very low mass objects must exist in
star forming regions, with infall rate and $\omega$ on average similar to
that of more massive cores.  Our conclusion that the spread of the measured
\macc\ is due to a comparable spread of the angular momentum at the onset of
the collapse and that cores of very different mass may rotate at the same
fraction of the break-up speed can also be tested by future 
observations.

Our models predict that the disk mass must depend strongly on the mass of
the central object (roughly as $\mstar^2$) and should be tightly
correlated with \macc. A significant number of disk mass measurements exist
only for the Taurus sample and are shown in Fig.~\ref{fig-mdisk}.  They have
been derived for TTS by Andrews and Williams (\citeyear{AW05}) by fitting
sub-mm and mm fluxes with simple disk models assuming a normalization of the
dust opacity of 10 cm$^2$ g$^{-1}$ at 300 $\mu$m, and a gas-to-dust mass
ratio of 100.  For BDs, we show the recent measurements of Scholz et
al.~(\citeyear{Sea06}), normalized to the same value of the opacity.  While
the distribution of the disk masses as a function of stellar mass does not
conflict with the model predictions, the expected correlation with \macc\ is
not seen in the data. Uncertainties in $\bcs$ and age will likely spread our
models, but will not undo the general predicted trend. Therefore there
appears to be a disagreement between the models and the observations in this
respect.  This may be an argument against our model.
However, measurements of disk masses
are severely affected by our ignorance
of the dust properties (Natta et
al.~\citeyear{NeaPPV}),  of the gas-to-dust ratio and of their
dependence on disk and stellar properties. In addition, in the BD
case one should note that the current millimeter detections are still very
uncertain and probably represent the outliers of the real distribution
(Scholz et al.~\citeyear{Sea06}).

The very simple models used in this Letter neglect several important aspects
of the star formation process, in particular the formation from a collapsing
core of multiple systems, rather than single stars.  A survey of disk
accretion rates of binary TTS in Taurus (White and Ghez \citeyear{WG01})
shows that \macc\ of the primaries does not differ from that of single
stars; Fig.~\ref{fig-main} includes a number of binaries, in which both
components are accreting, and we find that for both primary and secondary
\macc\ is well within the range of single stars with the same mass.  In the
context of our models, this is possible if the original core breaks into
fragments which evolve independently, with a distribution in $\omega$
similar to that of the cores from which single stars form.  However, we
point out that the majority of the objects in Fig.~\ref{fig-main} are
single, and for them the core-collapse models we use should be appropriate.

The similarity between the \macc--$\mstar$ properties of Taurus and Ophiuchus
originates, in our model, from a compensation between older age and lower
$\dminf$ for Taurus.  To disentangle these effects, one needs to measure
\macc\ in older star forming regions, where the age effect may become
important.  This will be challenging, as fewer stars should be accreting at
a detectable rate, and very low mass objects will very likely be below the
detection limit.  However, checking if the upper envelope of \macc\ is
consistent with our predictions will be important.

The model we propose explains the properties and evolution of disk accretion
entirely as a result of the initial condition for star formation and the
disk viscous evolution.  We neglect other processes, such as intermittent
accretion, photoevaporation, variability of the X-ray emission from the
central star, etc., which may affect at some level the snap-shot
measurements of \macc, but we claim that they do not control the basic
observed trends.

\section{Conclusion}\label{sec-conclusion}

In this Letter, we propose that the steep dependence of the mass accretion
rate on the mass of the central object, seen in two star forming regions of
different age and properties (Taurus and \Roph), results from the imprint of
the initial conditions for the formation of the star+disk system.  A simple
model of disk formation and evolution from collapsing cores naturally yields
the \macc $\propto \mstar^{1.8}$ relation, provided that cores of all mass
have a similar distribution of rotation rates in units of their break-up
speed. We show also that varying $\omega$ within a range consistent with the
(few) existing observations could account reasonably well for the large
spread of \macc\ observed at any $\mstar$, even though other factors may
also cause a spread. The predicted correlation of \macc\ with disk mass is
not well reproduced by the data; this requires further
investigations into the disk model and the disk mass measurements.

\begin{acknowledgements}
This project was partially supported by MIUR grant 2004025227/2004
to the INAF-Osservatorio di Arcetri.
\end{acknowledgements}


\begin{thebibliography}{4}
\expandafter\ifx\csname natexlab\endcsname\relax\def\natexlab#1{#1}\fi

\bibitem[Alexander \& Armitage (2006)]{AA06}
Alexander R.D., Armitage P.J. 2006, \apjl, 639, 83

\bibitem[Andrews \& Williams (2005)]{AW05}
Andrews, S.M., Williams, J.P. 2005, \apj, 631, 1134

\bibitem[Belloche et al.(2001)]{BAM01} Belloche, A., 
Andr{\'e}, P., \& Motte, F.\ 2001, ASP Conf.~Ser.~243: From Darkness to 
Light: Origin and Evolution of Young Stellar Clusters, 243, 313

\bibitem [Dullemond et al. (2006)]{DAW06}
Dullemond C.~P., Apai, D., Walch, S. 2006, \apjl, 640, 67

\bibitem[Goodman et al.(1993)]{Gea93} Goodman, A.~A., Benson, 
P.~J., Fuller, G.~A., \& Myers, P.~C.\ 1993, \apj, 406, 528 

\bibitem[Gullbring et al. (1998)]{Gea98}
Gullbring E., Hartmann L., Brice\~no C., Calvet N. 1998, \apj, 492, 323

\bibitem[Hartmann  et al. (1998)]{Hea98}
Hartmann, L., Calvet, N., Gullbring E., D'Alessio P., 1998, \apj, 495, 385

\bibitem[{{Hueso} \& {Guillot}(2005)}]{huesoguillot:2005}
{Hueso}, R. \& {Guillot}, T. 2005, \aap, 442, 703

\bibitem[Mohanty et al. (2005)]{Subu05}
Mohanty, S., Jayawardhana, R., Basri, G. 2005, \apj, 626, 498

\bibitem[Muzerolle et al. (2003)]{Mea03}
Muzerolle J., Hillenbrand L., Calvet N., Brice\~no C., Hartmann L. 2003,
\apj, 592, 266

\bibitem[Muzerolle et al. (2005)]{Mea05}
Muzerolle J., Luhman K.L., Brice\~no C., Hartmann L., Calvet N. 2005,
\apj, 625, 906

\bibitem[Myers \& Benson(1983)]{MB83} Myers, P.~C., \& 
Benson, P.~J.\ 1983, \apj, 266, 309

\bibitem[{{Nakamoto} \& {Nakagawa}(1995)}]{1995ApJ...445..330N}
{Nakamoto}, T. \& {Nakagawa}, Y. 1995, \apj, 445, 330

\bibitem[Natta et al. (2004)]{Nea04}
Natta A., Testi L., Muzerolle J., Randich S.,
 Comer\'on F., Persi P. 2004, \aap 424, 603

\bibitem[Natta et al. (2006a)]{Nea06a}
Natta A., Testi L., Randich S. 2006a, \aap, in press, astro-ph/0602618.

\bibitem [Natta et al. (2006b)]{NeaPPV}
Natta, A., Testi, L., Calvet, N., Henning, Th., Waters, R., Wilner, D.
2006b, in: {\it Protostars and Planets V}, in press. 

\bibitem [Padoan et al. (2005)]{Pea05}
Padoan, P., Kritsuk, A., Norman, M. L., Nordlund, Å.
2005, \apjl, 622, L61

\bibitem [Palla \& Stahler (2000)]{PS00}
Palla, F., Stahler, S. 2000, \apj, 540, 255

\bibitem[{Scholz et al. (2006)}]{Sea06}
Scholz, A., Jayawardhana, R., \& Wood, K., \apj, in press,  astro-ph/0603619

\bibitem[{Shakura \& Sunyaev(1973)}]{shaksuny:1973}
Shakura, N.~I. \& Sunyaev, R.~A. 1973, \aap, 24, 337

\bibitem[{{Shu}(1977)}]{Shu77}
{Shu}, F.~H. 1977, \apj, 214, 488

\bibitem[White \& Ghez (2001)]{WG01}
White R.J., Ghez A. 2001, ApJ 556, 265

\end{thebibliography}
\end{document}